\documentclass[prb,aps,twocolumn,floats,showpacs,preprintnumbers,amsmath,amssymb]{revtex4}

\usepackage{graphicx}
\usepackage{dcolumn}
\usepackage{bm}

\newcommand{\bq}{\begin{equation}}
\newcommand{\eq}{\end{equation}}
\newcommand{\bqa}{\begin{eqnarray}}
\newcommand{\eqa}{\end{eqnarray}}
\newcommand{\nn}{\nonumber \\}

\def\be     {\begin{equation}}
\def\ee     {\end{equation}}
\def\bea        {\begin{eqnarray}}
\def\eea        {\end{eqnarray}}
\def\bnn    {\begin{eqnarray*}}
\def\enn    {\end{eqnarray*}}

\begin{document}

\title{Critical particle-hole composites at twice the Fermi wave vector in U(1) spin liquid
with a Fermi surface}
\author{Ki-Seok Kim$^{1,2}$}
\affiliation{ $^1$Asia Pacific Center for Theoretical Physics,
Hogil Kim Memorial building 5th floor, POSTECH, Hyoja-dong, Namgu,
Pohang 790-784, Korea  \\ $^2$Department of Physics, Pohang
University of Science and Technology, Pohang, Gyeongbuk 790-784,
Korea }
\date{\today}

\begin{abstract}
We find "{\it chiral symmetry breaking}" at finite energies in
U(1) spin liquid, corresponding to critical particle-hole
composite states with twice of the Fermi momentum (2$k_{F}$). We
investigate this Fermi surface problem based on the
Nambu-Eliashberg theory, where the off diagonal pairing
self-energy is introduced to catch the Aslamasov-Larkin vertex
correction. This approach is quite parallel with the case of
superconductivity, where such Aslamasov-Larkin quantum corrections
in the particle-particle channel are well known to be responsible
for superconducting instability, formulated as the
Nambu-Eliashberg theory in an elegant way. We obtain the pairing
self-energy, which vanishes at zero energy but displays the same
power law dependence for frequency as the normal Eliashberg
self-energy. As a result, even the pairing self-energy correction
does not modify the Eliashberg dynamics without the Nambu spinor
representation, where thermodynamics is described by the typical
$z = 3$ scaling free energy. We discuss physical implication of
the anomalous self-energy identical to the conventional Eliashberg
normal self-energy, focusing on thermodynamics.
\end{abstract}


\maketitle

\section{Introduction}

Non-Fermi liquid physics in thermodynamics and transport has been
at the heart of condensed matter physics. Quantum criticality is
regarded as one source of such emergent phenomena \cite{HFQCP},
where exotica such as electron fractionalization
\cite{Lee_Nagaosa_Wen} and Landau-Ginzburg-Wilson forbidden
duality \cite{Senthil} beyond a microscopic model were proposed to
explain unique but universal critical phenomena.

An effective field theory for quantum critical dynamics can be
generally written in terms of renormalized electrons interacting
with order parameter fluctuations \cite{HMM}, associated with
either symmetry breaking or emergent gauge symmetry. Recently, it
has been clarified that two dimensional Fermi surface problems are
still strongly interacting even in the large-$N$ (spin degeneracy)
limit as the non-Abelian gauge theory
\cite{SungSik_Genus,Metlitski_Sachdev1,Metlitski_Sachdev2,McGreevy},
where one dimensional processes embedded in two dimensions, i.e.,
forward and backward scattering near the Fermi surface turn out to
be relevant \cite{Maslov_Review}. One patch formulation was
originally examined to keep only the forward scattering, where
Landau damping for critical boson dynamics is shown to be exact
\cite{SungSik_Genus}. Two patch construction was proposed to be a
minimal model for the Fermi surface problem, taking into account
not only the forward scattering but also the backscattering
associated with 2$k_{F}$ particle-hole excitations, where $k_{F}$
is the Fermi momentum. The Landau damping dynamics turns out to be
still preserved while the Eliashberg fermion self-energy gets
corrected via so called Aslamasov-Larkin (AL) vertex corrections,
perturbatively performed to result in an anomalous exponent
proportional to $1/N$ \cite{Metlitski_Sachdev1}. Critical boson
dynamics associated with SU(2) symmetry breaking (ferromagnetism)
was considered, arguing that even the dynamical exponent may be
modified \cite{Metlitski_Sachdev2}.

Unfortunately, all previous calculations have been performed in a
perturbative way based on the Eliashberg solution, thus it is
difficult to discuss on non-perturbative critical dynamics of
bosons and fermions. Recently, we performed an infinite order
summation for particular vertex corrections, given by ladder
diagrams \cite{Kim_Ladder}. This was performed in a fully
self-consistent way, resorting to the Ward identity. It turns out
that this class of quantum corrections does not modify the
Eliashberg dynamics in the one patch formulation.

AL vertex corrections were {\it perturbatively} demonstrated to
play an important role in both boson and fermion critical
dynamics, as mentioned before. In this respect we need to perform
an infinite order summation for this class of quantum corrections.
Physically speaking, these quantum processes are possible to cause
nontrivial dynamics for 2$k_{F}$ particle-hole excitations. It is
well known that the same class of quantum corrections is
responsible for superconducting instability of the Fermi liquid
state, where a pole with an imaginary frequency in the
particle-particle $t$-matrix is identified with a Cooper-pair
state \cite{BCS}. Similarly, 2$k_{F}$ particle-hole composites may
result from AL quantum processes, expected to modify the critical
dynamics.

In this paper we propose how to introduce the AL vertex correction
into the two dimensional U(1) gauge theory with a Fermi surface
{\it non-perturbatively}. Resorting to the analogy with
superconductivity, where the superconducting instability described
by the AL vertex correction is reformulated by the anomalous
self-energy in the Eliashberg framework of the Nambu spinor
representation \cite{BCS}, we argue that the off-diagonal
self-energy associated with the 2$k_{F}$ particle-hole channel
incorporates the same class of quantum corrections in the Nambu
spinor representation, where the Nambu spinor is composed of two
opposite Fermi momenta, $k_{F}$ and $-k_{F}$. We evaluate an
anomalous pairing self-energy in the Nambu-Eliashberg
approximation, which vanishes at zero energy but displays the same
power law dependence for frequency as the normal Eliashberg
self-energy. As a result, even the pairing self-energy correction
does not modify the Eliashberg dynamics without the Nambu spinor
representation, where thermodynamics is described by the typical
$z = 3$ scaling free energy. However, nonzero pairing self-energy
corrections at finite frequencies break "{\it chiral symmetry}" in
the two patch formulation of the U(1) gauge theory, indicating
critical particle-hole composite states with 2$k_{F}$. We discuss
physical implication of the anomalous self-energy identical to the
conventional Eliashberg normal self-energy, focusing on
thermodynamics.

\section{Nambu-Eliashberg theory for the Aslamasov-Larkin vertex correction}

\subsection{The ladder vertex correction}

We start from the two patch formulation for the two dimensional
U(1) gauge theory with a Fermi surface \bqa && {\cal S}_{eff} =
\int_{0}^{\beta} d \tau \int d^{2} r \Bigl\{ f_{s\sigma}^{\dagger}
\Bigl( \eta \partial_{\tau} - i s
\partial_{x} - \partial_{y}^{2} \Bigr) f_{s\sigma}
\nn && + \frac{s}{\sqrt{N}} a f_{s\sigma}^{\dagger} f_{s\sigma} +
a (- \partial_{y}^{2})^{\frac{z-1}{2}} a \Bigr\} , \eqa where
$f_{s\sigma}$ represents a spinon field and $a$ describes U(1)
gauge field in the spin liquid state. $s = \pm$ represents the
patch index, corresponding to $+ \equiv + k_{F}$ and $- \equiv -
k_{F}$, respectively, and $\sigma = 1, ..., N$ expresses the spin
index. Note that the sign in the gauge coupling is different
between the two patches, implying the gauge-current minimal
coupling. $\eta$ is an infinitesimal coefficient to control
artificial divergences in quantum corrections, which can be cured
by self-energy corrections \cite{SungSik_Genus}. $z$ is the
dynamical exponent determining the dispersion relation of gauge
fluctuations. It is given by $z = 3$ for several problems such as
ferromagnetic or nematic quantum criticality including the present
spin liquid problem \cite{Rech_Pepin_Chubukov} while $z = 2$ in
the spin density wave ordering \cite{HMM}. Both the Fermi velocity
$v_{F}$ and the curvature $1/m$ are set to one. Generally, one may
see that this effective field theory describes dynamics of
fermions interacting with collective bosons, associated with
symmetry breaking.

Dynamics of bosons and fermions can be described self-consistently
by their self-energy corrections. The fermion self-energy is
expressed as follows \begin{widetext} \bqa && \Sigma_{s}(k_0) = -
\frac{1}{N} \int \frac{d q_{0}}{2\pi} \int \frac{d^{2}
q}{(2\pi)^{2}} \Lambda(k_0+q_0,k+q;k_0,k) G_{s}(k_0+q_0,k+q)
D(q_0,q) \nn && - \frac{1}{N} \int \frac{d q_{0}}{2\pi} \int
\frac{d^{2} q}{(2\pi)^{2}} \Lambda_{2k_{F}}^{2}(k_0+q_0,k+q;k_0,k)
G_{-s}(k_0+q_0,k+q) D_{2k_{F}}(q_0, q) , \eqa
\end{widetext} where the first part results from the forward
scattering associated with $G_{s}(k_0+q_0,k+q)$ while the second
contribution comes from the backward interaction related with
$G_{-s}(k_0+q_0,k+q)$, not introduced in the one patch
formulation. $G_{s}(k_0,k)$ is the fully renormalized Green's
function and $D(q_0,q)$ is the fully renormalized boson
propagator, given by \bqa && G_{s} (k_0,k) = \frac{1}{i \eta k_{0}
+ s k_{x} + k_{y}^{2} - \Sigma_{s}(k_0)} , \nn && D(q_0,q) =
\frac{1}{|q_{y}|^{z-1} + \Pi(q_0,q)}  , \eqa respectively.
$\Pi(q_0,q)$ is the boson self-energy, basically given by the
fermion polarization function \begin{widetext} \bqa && \Pi(q_0,q)
= \int \frac{d k_{0}}{2\pi} \int \frac{d^{2} k}{(2\pi)^{2}}
\Lambda(k_0+q_0,k+q;k_0,k) G_{s}(k_0+q_0,k+q) G_{s}(k_0,k) , \nn
&& \Pi_{2k_{F}}(q_0, q) = \int \frac{d k_{0}}{2\pi} \int
\frac{d^{2} k}{(2\pi)^{2}} \Lambda_{2k_{F}}^{2}(k_0+q_0,k+q;k_0,k)
G_{s}(k_0+q_0,k+q) G_{-s}(k_0,k) , \eqa \end{widetext} where
$\Pi_{2k_{F}}(q_0, q)$ is the fermion polarization function around
the $2k_{F}$ transfer momentum.

\begin{figure}[t]
\includegraphics[width=0.40\textwidth]{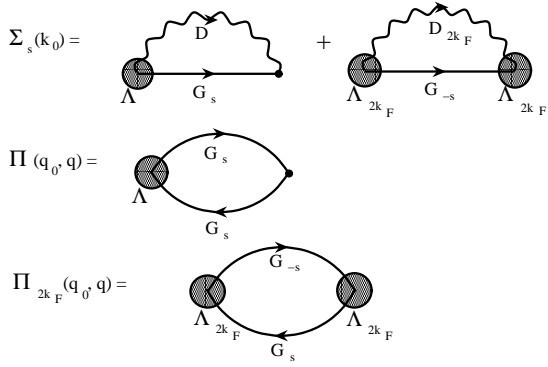}
\caption{Fermion self-energy [$\Sigma_{s}(k_{0})$] and boson
self-energies [$\Pi(q_0,q)$ and $\Pi_{2k_{F}}(q_0,q)$] : The
fermion self-energy results from both forward and backward
scattering and $2k_{F}$ polarization should be taken into account
explicitly in the boson self-energy. The thick line represents the
fermion Green's function and the wavy line does the gauge
propagator. The shaded region corresponds to the ladder-type
renormalized vertex.} \label{fig1}
\end{figure}

$\Lambda(k_0+q_0,k+q;k_0,k)$ and
$\Lambda_{2k_{F}}(k_0+q_0,k+q;k_0,k)$ are vertex corrections,
where the former is associated with the forward scattering and the
latter is related with the backscattering, given by
\begin{widetext} \bqa && \Lambda(k_0+q_0,k+q;k_0,k) = 1 \nn && - \int
\frac{d l_{0}}{2\pi} \int\frac{d^{2} l}{(2\pi)^{2}}
\Lambda(k_0+q_0-l_0,k+q-l; k_0-l_0,k-l) G_{s}(k_0-l_0,k-l)
G_{s}(k_0+q_0-l_0,k+q-l) D(l_0,l) \nn && - \int \frac{d
l_{0}}{2\pi} \int\frac{d^{2} l}{(2\pi)^{2}}
\Lambda(k_0+q_0-l_0-m_0,k+q-l-m ; k_0-l_0-m_0,k-l-m) \nn &&
G_{s}(k_0-l_0-m_0, k-l-m) G_{s}(k_0+q_0-l_0-m_0, k+q-l-m)
D_{2k_{F}}(m_0,m) \nn && G_{-s}(k_0-l_0,k-l)
G_{s}(k_0+q_0-l_0,k+q-l) D_{2k_{F}}(l_0,l)    \eqa and \bqa &&
\Lambda_{2k_{F}}(k_0+q_0,k+q;k_0,k) = 1 \nn && - \int \frac{d
l_{0}}{2\pi} \int\frac{d^{2} l}{(2\pi)^{2}}
\Lambda_{2k_{F}}(k_0+q_0-l_0,k+q-l; k_0-l_0,k-l)
G_{-s}(k_0-l_0,k-l) G_{s}(k_0+q_0-l_0,k+q-l) D(l_0,l) \nn && -
\int \frac{d l_{0}}{2\pi} \int\frac{d^{2} l}{(2\pi)^{2}}
\Lambda_{2k_{F}}(k_0+q_0-l_0-m_0,k+q-l-m; k_0-l_0-m_0,k-l-m) \nn
&& G_{-s}(k_0-l_0-m_0, k-l-m) G_{s}(k_0+q_0-l_0-m_0, k+q-l-m)
D_{2k_{F}}(m_0,m) \nn && G_{s}(k_0-l_0,k-l)
G_{s}(k_0+q_0-l_0,k+q-l) D_{2k_{F}}(l_0,l) , \eqa \end{widetext}
respectively, in the ladder approximation. It is straightforward
to read these vertex corrections. The first part in the forward
scattering vertex is that $G_{s}(k_0,k)$ emits $D(l_0,l)$ to be
$G_{s}(k_0-l_0,k-l)$ and scatters into $G_{s}(k_0+q_0-l_0,k+q+l)$
by $\Lambda(k_0+q_0-l_0,k+q-l; k_0-l_0,k-l)$, accepting $D(l_0,l)$
to be $G_{s}(k_0+q_0,k+q)$. In the second correction
$G_{s}(k_0,k)$ emits $D_{2k_{F}}(l_0,l)$ to be
$G_{-s}(k_0-l_0,k-l)$, which scatters into
$G_{s}(k_0-l_0-m_0,k-l-m)$ by $D_{2k_{F}}(m_0,m)$.
$G_{s}(k_0-l_0-m_0,k-l-m)$ becomes $G_{s}(k_0+q_0-l_0-m_0,
k+q-l-m)$ by $\Lambda(k_0+q_0-l_0-m_0,k+q-l-m ;
k_0-l_0-m_0,k-l-m)$, and it accepts $D_{2k_{F}}(m_0,m)$ to be
$G_{s}(k_0+q_0-l_0,k+q-l)$. It again accepts $D_{2k_{F}}(l_0,l)$
to be $G_{s}(k_0+q_0,k+q)$. One can understand the 2$k_{F}$ vertex
correction in the same way. Equations (2), (3), (4), (5), and (6)
consist of a fully self-consistent framework for a particular
class of quantum corrections in the Fermi surface problem.

\begin{figure}[t]
\includegraphics[width=0.40\textwidth]{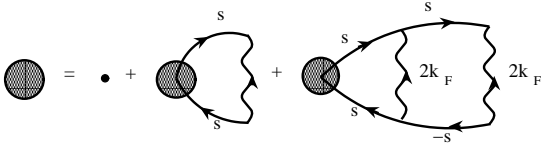}
\caption{(Color online) The ladder vertex correction near the zero
momentum transfer turns out to be irrelevant in the Eliashberg
solution.} \label{fig2}
\end{figure}

\begin{figure}[t]
\includegraphics[width=0.40\textwidth]{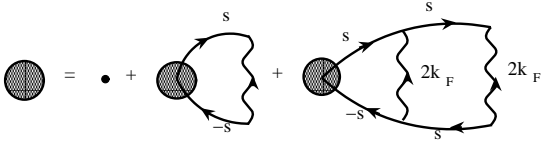}
\caption{(Color online) The ladder vertex correction near the
$2k_{F}$ transfer momentum turns out to be singular in the
Eliashberg solution, and should be taken into account carefully.}
\label{fig3}
\end{figure}

If the backward scattering given by
$\Lambda_{2k_{F}}(k_0+q_0,k+q;k_0,k)$ is neglected, these
self-consistent equations are reduced to those of the one patch
formulation. As mentioned in the introduction, such forward
scattering contributions do not modify the Eliashberg dynamics
\cite{Kim_Ladder}. The ladder vertex correction was solved,
resorting to the Ward identity with an ansatz, where the vertex
function is given by the fully renormalized Green's function. Such
an ansatz turns out to recover both the non-interacting limit and
one dimensional case. Inserting this ansatz expression into both
self-energy equations, we obtain fully self-consistent solutions,
basically the same as the Eliashberg self-energies in the low
energy limit.

However, the role of the backscattering was emphasized because the
$2k_{F}$ ladder vertex turns out to be singular \cite{U1SL_N},
given by \bqa \Lambda_{2k_{F}}(k_0+q_0,k+q;k_0,k) =
\frac{\Lambda_{0}}{\Bigl\{ \Bigl|\frac{q_{0}}{\epsilon_{F}}\Bigr|
+ c \Bigl( \frac{\sqrt{q_{x}^{2}+q_{y}^{2}}}{2k_{F}}
\Bigr)^{\frac{3}{2}} \Bigr\}^{\eta}} , \eqa where the transfer
momentum $q$ is already expanded near $2k_{F}$. $\epsilon_{F}$ is
the Fermi energy and $c$ is a positive numerical constant of the
order of one. $\Lambda_{0}$ is the bare vertex, set to be $1$ in
this paper. It should be addressed that this singular $2k_{F}$
particle-hole ladder vertex was not obtained self-consistently.
Instead, it was found from a perturbative summation up to an
infinite order (parquet approximation), based on the Eliashberg
Green's function \cite{U1SL_N}. Unfortunately, we do not know how
to get a self-consistent vertex solution, where the Ward identity
cannot be used for the $2k_{F}$ channel. In this subsection we
discuss several possibilities based on the previous perturbative
vertex solution.

$\eta$ is an anomalous exponent for the $2k_{F}$ vertex function,
depending on the spin degeneracy and given by $1/N$ basically.
Inserting this vertex into Eq. (4), the previous analysis found
two kinds of boson self-energy corrections \cite{U1SL_N} \bqa &&
\Pi_{2k_{F}}(q_0, q; \eta < \eta_{c}) \nn && = \mathcal{P} -
\sqrt{\frac{p_{0}}{\omega_{0}v_{F}^{3}}} \Bigl\{ c_{\omega} \Bigl(
\frac{q_{0}}{\omega_{0}} \Bigr)^{\frac{2}{3} - 2 \eta} + c_{q}
\Bigl| \frac{\sqrt{q_{x}^{2}+q_{y}^{2}}}{2k_{F}} \Bigr|^{1 - 3
\eta} \Bigr\} , \nn && \Pi_{2k_{F}}(q_0, q; \eta
> \eta_{c}) \nn && = \sqrt{\frac{p_{0}}{\omega_{0}v_{F}^{3}}} \Bigl\{
c_{\omega} \Bigl( \frac{q_{0}}{\omega_{0}} \Bigr)^{2 \eta -
\frac{2}{3} } + c_{q} \Bigl|
\frac{\sqrt{q_{x}^{2}+q_{y}^{2}}}{2k_{F}} \Bigr|^{ 3 \eta - 1 }
\Bigr\}^{-1} , \eqa where the critical value is $\eta_{c} = 1/3$.
$\omega_{0} = \Bigl( \frac{1}{2\sqrt{3}}\Bigr)^{3}
\frac{2v_{F}^{3}g^{4}}{\pi^{2} p_{0}}$, where $v_{F}$ and $p_{0}$
are the Fermi velocity and curvature, respectively, and $g$ is the
gauge coupling constant. Remember $v_{F}, p_{0}, g = 1$ in this
paper.

The 2$k_{F}$ boson self-energy does not diverge in $\eta <
\eta_{c}$ and can be neglected at low energies due to a positive
constant $\mathcal{P}$. On the other hand, it becomes singular
when $\eta$ exceeds $\eta_{c}$, implying quantum criticality of
the boson state. An important question is how the fermion
self-energy behaves, scattered by this 2$k_{F}$ critical boson
dynamics. Does it diverge to give rise to an instability to a
density wave (charge or spin) or vanish with an anomalous
exponent? The latter will be identified with a quantum critical
state, characterized by a diverging 2$k_{F}$ spin susceptibility.
We believe that this question is not addressed appropriately,
i.e., fully self-consistently even in the relativistic U(1) gauge
theory, where the $1/N$ expansion is well defined. It was
demonstrated that the ladder vertex correction results in
singularity to the staggered spin susceptibility, based on the
$1/N$ expansion in the relativistic U(1) gauge theory
\cite{Wen_ASL}. Since the fermion self-energy vanishes with an
anomalous critical exponent, the quantum criticality characterized
by the divergent staggered spin susceptibility can be regarded as
a stable phase, sometimes identified with an algebraic spin liquid
as one possible phase of the paramagnetic Mott insulator
\cite{Lee_Nagaosa_Wen}. This statement is certainly plausible at
$N = N_{c}$, below which chiral symmetry breaking arises to result
in an antiferromagnetic order, because $N = N_{c}$ corresponds to
an antiferromagnetic quantum critical point \cite{Don_Kim}. Our
real question is whether this critical solution survives or not
away from the quantum critical point, i.e., in the case when $N >
N_{c}$. Although the previous non-self-consistent analysis
demonstrated quantum criticality of the antiferromagnetic spin
susceptibility in the algebraic spin liquid state ($N > N_{c}$),
it seems to be possible that a fully self-consistent treatment may
cause both the fermion self-energy and the staggered spin
susceptibility to vanish with non-trivial critical exponents.

The present Fermi surface problem is much more complicated because
the $1/N$ expansion is not well defined \cite{SungSik_Genus}.
Inserting the boson self-energy of Eq. (8) into Eq. (2), one can
find the fermion self-energy, particulary corrected by the
$2k_{F}$ process. As discussed above, it is not clear whether
$\eta$ exceeds its critical value or not. When $\eta$ is smaller
than $\eta_{c}$, the $2k_{F}$ scattering is irrelevant due to
gapped boson excitations at this momentum, and the Eliashberg
dynamics will be preserved. However, the divergent boson
self-energy in the case of $\eta > \eta_{c}$ makes the fermion
self-energy more singular than the Eliashberg one. Actually,
inserting the singular $2k_{F}$ ladder vertex and the divergent
$2k_{F}$ boson self-energy into the equation for the fermion
self-energy, we obtain \bqa && \Sigma_{2k_{F}}(k_0; \eta
> \eta_{c}) \propto - i \frac{\mbox{sgn}(k_{0})}{N}
|k_{0}|^{\frac{1}{3} - 2 \eta} , \eqa which diverges in the $k_{0}
\rightarrow 0$ limit. This can be interpreted as a particle-hole
bound state, analogous with a Cooper pair in the particle-particle
channel, resulting in a density wave state, where gauge
fluctuations play the role of a pairing glue, reflected in their
divergent self-energy.

We believe that this 2$k_{F}$ issue should be addressed
self-consistently in both relativistic and non-relativistic U(1)
gauge theories. One way to resolve this issue is discussed in the
last subsection. In this paper we assume that the $2k_{F}$ ladder
process does not cause any change to the Eliashberg dynamics,
i.e., $\eta < \eta_{c}$, and investigate other quantum
corrections.

\subsection{Review on the Aslamasov-Larkin vertex
correction associated with the superconducting instability}

We review the role of AL quantum corrections in the Eliashberg
dynamics, demonstrated to cause an anomalous exponent beyond the
Eliashberg fermion self-energy. As discussed in the introduction,
such AL vertex corrections in the particle-particle channel were
shown to cause superconducting instability to the Fermi liquid
state \cite{BCS}. Interestingly, this $t$-matrix construction was
reformulated in the particle-hole Nambu representation, resulting
in the same critical temperature. In this subsection we focus on
the connection between the Nambu-Eliashberg theory and the AL
quantum corrected self-energy.

The electron self-energy is given by \begin{widetext} \bqa &&
\Sigma (p_0) = - \int \frac{d q_{0}}{2\pi} \int \frac{d^{3}
q}{(2\pi)^{3}} \Gamma (p_0+q_0,p+q;p_0,p) G (p_0+q_0,p+q) g_{q} D
(q_0,q) , \eqa \end{widetext} where $g_{q}$ is the electron-phonon
coupling constant and $\Gamma (p_0+q_0,p+q;p_0,p)$ is the
associated vertex. $D (q_0,q)$ is the phonon propagator. Here, the
imaginary time formulation is applied at zero temperature. The
electron-phonon vertex can be approximated as follows (Fig. 4)
\begin{widetext} \bqa \Gamma (p_0+q_0,p+q;p_0,p) &=& g_{q} \nn &-&
g_{q} \int \frac{d k_0}{ 2\pi }  \int \frac{d^{3} k}{(2\pi)^{3}}
\mathbf{T}_{pp}[(k_0,k;p_0+q_0,p+q)(k_0+q_0,k+q;p_0,p)]
G(k_0+q_0,k+q) G(k_0,k) , \nn \eqa \end{widetext} where
$\mathbf{T}_{pp}[(k_0,k;p_0+q_0,p+q)(k_0+q_0,k+q;p_0,p)]$ is the
particle-particle $t$-matrix to show the transition process that
$(p_0,p)$ and $(k_0+q_0,k+q)$ scatter into $(p_0+q_0,p+q)$ and
$(k_0,k)$ with the transfer momentum $(q_0,q)$, respectively.
Inserting this AL vertex into the fermion self-energy, we obtain
\begin{widetext}
\bqa && \Sigma (p_0) \approx - \int \frac{d q_{0}}{2\pi} \int
\frac{d^{3} q}{(2\pi)^{3}} G (p_0+q_0,p+q) g_{q} D (q_0,q) g_{q}
\nn && + \int \frac{d q_{0}}{2\pi} \int \frac{d^{3} q}{(2\pi)^{3}}
\mathbf{T}_{pp}[(-p_0,-p;p_0+q_0,p+q)(-p_0+q_0,-p+q;p_0,p)] \nn &&
G (p_0+q_0,p+q) G(-p_0,-p) G(-p_0+q_0,-p+q) g_{q} D (q_0,q) g_{q}
, \eqa \end{widetext} where the most singular particle-particle
channel was selected (Fig. 5).

\begin{figure}[t]
\includegraphics[width=0.40\textwidth]{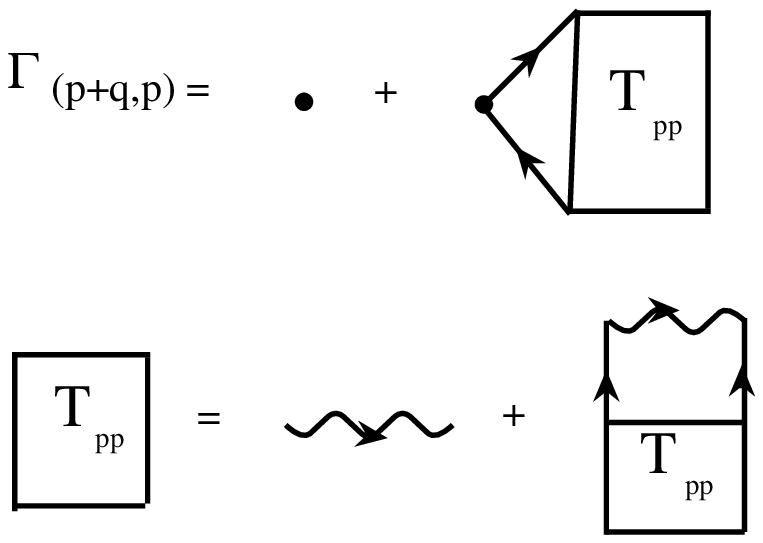}
\caption{(Color online) The AL vertex correction in the
particle-particle channel for superconducting instability}
\label{fig4}
\end{figure}

\begin{figure}[t]
\includegraphics[width=0.32\textwidth]{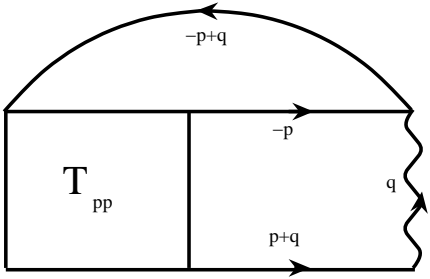}
\caption{(Color online) The electron self-energy by the AL vertex
correction in the particle-particle channel for superconducting
instability} \label{fig5}
\end{figure}

The particle-particle $t$-matrix is given by the Bethe-Salpeter
equation \cite{BCS} \begin{widetext} \bqa &&
\mathbf{T}_{pp}[(-p_0,-p;p_0+q_0,p+q)(-p_0+q_0,-p+q;p_0,p)] =
g_{q} D (q_0,q) g_{q} \nn && + \int \frac{d k_0'}{ 2\pi }  \int
\frac{d^{3} k'}{(2\pi)^{3}}
\mathbf{T}_{pp}[(-p_0+q_0-k_0',-p+q-k';p_0+k_0',p+k')(-p_0+q_0,-p+q;p_0,p)]
\nn && G(p_0+k_0',p+k') G(-p_0+q_0-k_0',-p+q-k') g_{q-k'} D
(q_0-k_0',q-k') g_{q-k'} .  \eqa \end{widetext} See Fig. 4.
Generally speaking, this integral equation cannot be solved
analytically because it is not factorized. But, if the retardation
effect can be neglected, one obtains its analytic expression
\cite{BCS}.

We compare this approach with an elegant reformulation based on
the Nambu-Eliashberg approximation. The Dyson equation is \bqa &&
\boldsymbol{G}(p_0,p) = \boldsymbol{g}(p_0,p) +
\boldsymbol{g}(p_0,p) \boldsymbol{\Sigma}(p_0)
\boldsymbol{G}(p_0,p) , \nonumber \eqa where the Green's function
is a two by two matrix in the Nambu spinor representation, given
by \bqa && \boldsymbol{G}(p_0,p) \equiv - \Bigl\langle  \left(
\begin{array}{c} c_{p\uparrow} \\
c_{-p\downarrow}^{\dagger} \end{array} \right)  \left(
\begin{array}{cc} c_{p\uparrow}^{\dagger} &
c_{-p\downarrow} \end{array} \right) \Bigr\rangle \nn && \equiv
\left(
\begin{array}{cc} G(p_0,p) & F(p_0,p) \\
F^{*}(p_0,p) & - G(-p_0,-p) \end{array} \right) . \nonumber \eqa
$G(p_0,p)$ is a normal Green's function with a bare propagator
$g(p_0,p)$ while $F(p_0,p)$ is an anomalous propagator associated
with particle-particle pairing. The electron self-energy is given
by
\bqa && \boldsymbol{\Sigma}(p_0) = \left( \begin{array}{cc} {\Sigma}(p_{0}) & \Phi(p_{0}) \\
\Phi^{*}(p_{0}) & - {\Sigma} (-p_{0}) \end{array} \right) ,
\nonumber \eqa where $\Phi(p_{0})$ is the pairing self-energy and
$\Sigma(p_{0})$, the normal one.  Then, the Dyson equation becomes
\begin{widetext}
\bqa && G(p_0,p) = g(p_0,p) + g(p_0,p) {\Sigma}(k_{0})G(p_0,p) +
g(p_0,p) \Phi(p_{0}) F^{*}(p_0,p) , \nn && F^{*}(p_0,p) = -
g(-p_0,-p) \Phi^{*}(p_{0}) G(p_0,p) + g(-p_0,-p) {\Sigma} (-p_{0})
F^{*}(p_0,p) . \eqa \end{widetext} Inserting  \bqa && F^{*}(p_0,p)
= - \frac{g (-p_0,-p) \Phi^{*}(p_{0}) G(p_0,p)}{1 - g (-p_0,-p)
{\Sigma} (-p_{0})} \nonumber \eqa from the second equation into
the first, we obtain the full renormalized self-energy in the
Nambu formalism \bqa && \Sigma_{tot}(p_0) = \Sigma(p_0) - \frac{g
(-p_0,-p) |\Phi(p_{0})|^{2}}{1 - g (-p_0,-p) {\Sigma} (-p_{0})} ,
\eqa where the total self-energy is defined as \bqa && G(p_0,p)
\equiv g(p_0,p) + g(p_0,p) \Sigma_{tot}(p_0) G(p_0,p) . \nonumber
\eqa

The self-energy is determined in the Eliashberg approximation \bqa
&& \boldsymbol{\Sigma}(p_0) = \int\frac{d q_0}{2\pi} \int
\frac{d^{3} q}{(2\pi)^{3}} g_{ph}^{2} D(q_0,q)
\boldsymbol{G}(p_0+q_0,p+q) . \nn \eqa Inserting the Nambu Green's
function into the above expression, we find full self-consistent
equations
\begin{widetext}
\bqa && \Phi^{*}(p_0) = - \int\frac{d q_0}{2\pi} \int \frac{d^{3}
q}{(2\pi)^{3}} g_{ph}^{2} D(q_0,q) \frac{g(-p_0-q_0,-p-q)
\Phi^{*}(p_0+q_0) }{1 - g(-p_0-q_0,-p-q) {\Sigma} (-p_0-q_0)} \nn
&& \frac{g(p_0+q_0,p+q)}{1 - g(p_0+q_0,p+q) \Bigl\{ \Sigma(p_0+q_0
) - \frac{g(-p_0-q_0,-p-q) |\Phi(p_{0}+q_0)|^{2}}{1 -
g(-p_0-q_0,-p-q) {\Sigma} (-p_{0}-q_0)} \Bigr\}} \eqa for the
pairing self-energy and \bqa && {\Sigma}(p_0) = \int\frac{d
q_0}{2\pi} \int \frac{d^{3} q}{(2\pi)^{3}} g_{ph}^{2} D(q_0,q)
\frac{g(p_0+q_0,p+q)}{1 - g(p_0+q_0,p+q) \Bigl\{ \Sigma(p_0+q_0 )
- \frac{g(-p_0-q_0,-p-q) |\Phi(p_{0}+q_0)|^{2}}{1 -
g(-p_0-q_0,-p-q) {\Sigma} (-p_{0}-q_0)} \Bigr\}} \eqa
\end{widetext} for the normal electron self-energy.

In order to see similarity between the $t$-matrix approximation
and the Nambu-Eliashberg one, we perform the following
approximation for the normal self-energy \begin{widetext} \bqa &&
{\Sigma}(p_0) \approx \int\frac{d q_0}{2\pi} \int \frac{d^{3}
q}{(2\pi)^{3}} g(p_0+q_0,p+q) g_{ph} D(q_0,q) g_{ph} \nn && -
\int\frac{d q_0}{2\pi} \int \frac{d^{3} q}{(2\pi)^{3}}
|\Phi(p_{0}+q_0)|^{2} g(p_0+q_0,p+q) g(p_0+q_0,p+q)
g(-p_0-q_0,-p-q) g_{ph} D(q_0,q) g_{ph} . \eqa
\end{widetext} Comparing this expression with Eq. (12) in
the $t$-matrix self-energy, we find some similarity if we identify
the $t$-matrix [Eq. (13)] with the square of the pairing
self-energy [Eq. (17)] in the following approximation
\begin{widetext} \bqa && \Phi^{*}(p_0) \approx - \int\frac{d
q_0}{2\pi} \int \frac{d^{3} q}{(2\pi)^{3}} g_{ph}^{2} D(q_0,q)
g(p_0+q_0,p+q)g(-p_0-q_0,-p-q) \Phi^{*}(p_0+q_0) . \eqa
\end{widetext}

It is clear that the $t$-matrix approximation is not equivalent to
the Nambu-Eliashberg formulation in the normal state because the
anomalous pairing self-energy should vanish as discussed in the
next section. However, we believe that the anomalous self-energy
in the Nambu-Eliashberg theory will catch an essential physics of
the AL vertex-corrected electron self-energy at the critical
point, where the singular part of the normal self-energy results
from the pairing one in the Nambu-Eliashberg approximation while
from the $t$-matrix process in the AL vertex correction. One can
regard the Nambu representation as one possible way to select a
better basis set for an instability channel via "unitary
transformation".

\subsection{Nambu-Eliashberg theory}

One can construct a self-consistent framework for boson and
fermion self-energy corrections, introducing the AL vertex with
the 2$k_{F}$ particle-hole $t$-matrix Bethe-Salpeter equation.
However, this strategy is not practical because such
self-consistent equations are much complicated, where the
retardation effect is inevitable. As the AL vertex correction is
reformulated based on the Nambu spinor representation in
superconductivity, we introduce such quantum processes in the same
way, but the component of the Nambu spinor will be different from
that in superconductivity.

Resorting to the analogy with superconductivity, we rewrite the
two patch model Eq. (1) in the Nambu spinor representation \bqa &&
{\cal S}_{eff} = \int_{0}^{\beta} d \tau \int d^{2} r \Bigl\{
F_{\sigma}^{\dagger} \Bigl( \eta
\partial_{\tau} \boldsymbol{I} - i
\partial_{x} \boldsymbol{\tau_{3}} - \partial_{y}^{2} \boldsymbol{I}
\Bigr) F_{\sigma} \nn && + \frac{1}{\sqrt{N}} a
F_{\sigma}^{\dagger} \boldsymbol{\tau_{3}}  F_{\sigma} + a (-
\partial_{y}^{2})^{\frac{z-1}{2}} a \Bigr\} , \eqa where
$F_{\sigma} = \left( \begin{array}{c} f_{+\sigma} \\
f_{-\sigma} \end{array} \right)$ is the Nambu spinor composed of
each-patch fermion. $\boldsymbol{\tau_{3}}$ acts on the $+ k_{F}$
particle and $- k_{F}$ particle space.

Both boson and fermion self-energies can be found in the
Nambu-Eliashberg theory \begin{widetext} \bqa && \Pi(q_0,q) =
\boldsymbol{tr} \int \frac{d k_{0}}{2\pi} \int \frac{d^{2}
k}{(2\pi)^{2}} \boldsymbol{\tau_{3}} \boldsymbol{G}(k_0+q_0,k+q)
\boldsymbol{\tau_{3}} \boldsymbol{G}(k_0,k) , \nn &&
\boldsymbol{\Sigma}(k_0) = - \frac{1}{N} \int \frac{d q_{0}}{2\pi}
\int \frac{d^{2} q}{(2\pi)^{2}} \boldsymbol{\tau_{3}}
\boldsymbol{G}(k_0+q_0,k+q) D(q_0,q) ,  \eqa \end{widetext} where
\bqa \boldsymbol{G}(k_0,k) &=& \Bigl( i \eta k_{0} \boldsymbol{I}
+ k_{x} \boldsymbol{\tau_{3}} + k_{y}^{2} \boldsymbol{I} -
\boldsymbol{\Sigma}(k_{0}) \Bigr)^{-1} \nonumber  \eqa is the
fermion Nambu Green's function with the matrix self-energy \bqa &&
\boldsymbol{\Sigma}(k_{0}) =
\left( \begin{array}{cc} {\Sigma}(k_{0}) & \Phi(k_{0}) \\
\Phi^{*}(k_{0}) & {\Sigma}(k_{0}) \end{array} \right) . \nonumber
\eqa ${\Sigma}(k_{0})$ is the normal part and $\Phi(k_{0})$ is an
effective pairing potential. More explicitly, the normal fermion
self-energy, the fermion pairing potential, and the boson
self-energy are given by \begin{widetext} \bqa {\Sigma}(k_0) &=& -
\frac{1}{N} \int \frac{d q_{0}}{2\pi} \int \frac{d^{2}
q}{(2\pi)^{2}} G(k_0+q_0,k+q) D(q_0,q) , \nn \Phi(k_{0}) &=&
\frac{1}{N} \int \frac{d q_{0}}{2\pi} \int \frac{d^{2}
q}{(2\pi)^{2}} F(k_0+q_0,k+q) D(q_0,q)  , \nn  \Pi(q_0,q) &=& \int
\frac{d k_{0}}{2\pi} \int \frac{d^{2} k}{(2\pi)^{2}} \Bigl\{
G(k_0+q_0,k+q)G(k_0,k) + G(k_0+q_0,-k-q)G(k_0,-k) \nn &-&
F(k_0+q_0,k+q) F^{*}(k_0,k) - H.c. \Bigr\} , \eqa where \bqa &&
G(k_0,k) = \frac{i \eta k_{0} - k_{x} + k_{y}^{2} -
{\Sigma}(k_{0})}{\Bigl(i \eta k_{0} + k_{y}^{2} -
{\Sigma}(k_{0})\Bigr)^{2} - k_{x}^{2} - |\Phi(k_{0})|^{2}} , \nn
&& F(k_0,k) = - \frac{\Phi(k_{0})}{\Bigl(i \eta k_{0} + k_{y}^{2}
- {\Sigma}(k_{0})\Bigr)^{2} - k_{x}^{2} - |\Phi(k_{0})|^{2}} \eqa
\end{widetext} are normal and abnormal fermion propagators, respectively.

\subsection{A solution of the Eliashberg equation}

Although it is not easy to solve Eq. (23) in a general way, the
Landau damping ansatz for the boson self-energy \bqa && \Pi(q_0,q)
= \gamma \frac{|q_0|}{|q_y|} \eqa makes this problem
straightforward. In appendix we will check that the Landau damping
boson self-energy is self-consistent indeed.

Inserting the Landau damping boson self-energy into the boson
propagator, normal and anomalous fermion self-energies satisfy
\begin{widetext}  \bqa && {\Sigma}(k_0) = \frac{i}{2N} \Bigl\{ \int_{0}^{\infty}
\frac{d y}{ \pi } \frac{ y }{ y^{z} + 1 } \Bigr\}
\int_{-k_{0}}^{0} \frac{d q_{0}}{2\pi} (\gamma
|q_{0}|)^{\frac{2-z}{z}} \frac{ {\Sigma}(k_{0}+q_{0})}{ \sqrt{
{\Sigma}^{2}(k_{0}+q_{0}) - |\Phi(k_{0}+q_{0})|^{2}}} , \nn &&
\Phi(k_{0}) = \frac{i}{2N} \Bigl\{ \int_{0}^{\infty} \frac{d y}{
\pi } \frac{ y }{ y^{z} + 1 } \Bigr\} \int_{-k_{0}}^{0} \frac{d
q_{0}}{2\pi} (\gamma |q_{0}|)^{\frac{2-z}{z}}
\frac{\Phi(k_{0}+q_{0})}{ \sqrt{ {\Sigma}^{2}(k_{0}+q_{0}) -
|\Phi(k_{0}+q_{0})|^{2}}} . \eqa \end{widetext}  These
self-consistent equations are typical in the Nambu-Eliashberg
theory.

Considering the symmetry ${\Sigma}(k_0) \longleftrightarrow
\Phi(k_{0})$ in Eq. (26), we propose \bqa && \Sigma(k_{0}) = - i
\frac{\lambda }{N} \mbox{sgn}(k_0) |k_0|^{\frac{2}{z}} , \nn &&
\Phi(k_{0}) = - i \frac{\chi}{N} \mbox{sgn}(k_0)
|k_0|^{\frac{2}{z}} \eqa as a possible solution. Inserting this
expression into Eq. (26), we see
\begin{widetext} \bqa && {\Sigma}(k_0) = - \frac{i}{2N} \gamma^{\frac{2-z}{z}}
\Bigl\{ \int_{0}^{\infty} \frac{d y}{ \pi } \frac{ y }{ y^{z} + 1
} \Bigr\} \Bigl\{ \int_{-1}^{0} \frac{d y}{2\pi}
|y|^{\frac{2-z}{z}} \Bigr\} \mbox{sgn}(k_0) |k_0|^{\frac{2}{z}}
\frac{ \lambda }{ \sqrt{ \lambda^{2} + \chi^{2} }}
, \nn && \Phi(k_{0}) = - \frac{i}{2N} \gamma^{\frac{2-z}{z}}
\Bigl\{ \int_{0}^{\infty} \frac{d y}{ \pi } \frac{ y }{ y^{z} + 1
} \Bigr\} \Bigl\{ \int_{-1}^{0} \frac{d y}{2\pi}
|y|^{\frac{2-z}{z}} \Bigr\} \mbox{sgn}(k_0) |k_0|^{\frac{2}{z}}
\frac{ \chi }{ \sqrt{ \lambda^{2} + \chi^{2} }}
. \nonumber \eqa \end{widetext} Equating this with Eq. (27), we
obtain \bqa && \sqrt{ \lambda^{2} + \chi^{2} } =
\frac{\gamma^{\frac{2-z}{z}} }{2} \Bigl\{ \int_{0}^{\infty}
\frac{d y}{ \pi } \frac{ y }{ y^{z} + 1 } \Bigr\} \Bigl\{
\int_{-1}^{0} \frac{d y}{2\pi} |y|^{\frac{2-z}{z}} \Bigr\} , \nn
\eqa where $\lambda$ and $\chi$ cannot be determined, but only
$\lambda^{2} + \chi^{2}$ is given by Eq. (28). This is the trace
of the ${\Sigma}(k_0) \longleftrightarrow \Phi(k_{0})$ symmetry in
Eq. (26).

Appearance of the off diagonal self-energy is far from triviality.
If the self-consistent equation for the anomalous part is
linearized, one obtains the matrix equation \bqa &&
\boldsymbol{A}_{k_{0}} = \boldsymbol{M}_{k_{0}k_{0}'}
\boldsymbol{A}_{k_{0}'} , \nonumber \eqa where
$\boldsymbol{A}_{k_{0}}$ corresponds to the pairing self-energy
and the column index is frequency from zero to cutoff. This
homogeneous equation does not have a nontrivial solution generally
except the case when the matrix $\boldsymbol{M}_{k_{0}k_{0}'}$
satisfies the particular condition of $\mathbf{Det}
\bigl(\boldsymbol{M} - p \boldsymbol{I} \bigr) = 0$, where $p$ is
the eigen value set to be $1$. Such a condition can be satisfied
only at $T \leq T_{c}$ in the case of superconductivity, and the
corresponding eigen vector $\boldsymbol{A}_{k_{0}}$ vanishes
identically when $T > T_{c}$. The present situation is quite
analogous with the $T = T_{c}$ case, i.e., the critical point,
where $\boldsymbol{A}_{k_{0} \rightarrow 0} \rightarrow 0$ but
$\boldsymbol{A}_{k_{0} \not= 0} \not= 0$. The off diagonal
self-energy correction may be regarded as one characteristic
feature of the spin liquid state with a Fermi surface.

An interesting result from the anomalous self-energy is that the
chiral symmetry $\boldsymbol{U} = \exp(i \theta
\boldsymbol{\tau}_{3})$ is broken at nonzero frequencies,
indicating critical particle-hole composites with $2k_{F}$. Figure
6 shows an imaginary part of the anomalous propagator in Eq. (24).
As frequency increases, it is enhanced to a certain frequency and
decreased, following the self-energy behavior. The maximum point
becomes larger and its width gets broadened as the momentum
$k_{x}$ of fermions increases. Since the anomalous pairing
self-energy is identical to the normal self-energy, the normal
Green's function displays qualitatively the same feature as the
anomalous propagator. Appearance of critical particle-hole
composites with $2k_{F}$ is a novel feature of the two dimensional
U(1) gauge theory with a Fermi surface.

\begin{figure}[t]
\includegraphics[width=0.32\textwidth]{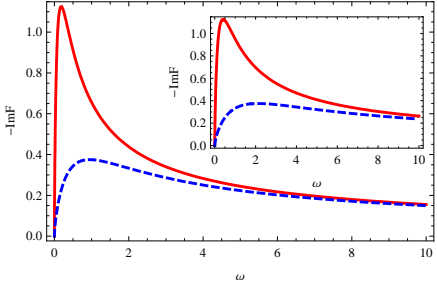}
\caption{(Color online) An imaginary part of the anomalous
propagator in Eq. (24). The thick (red) line corresponds to $k_{x}
= 0.1$ and $k_{y} = 0.0$ while the dashed (blue) line does to
$k_{x} = 0.3$ and $k_{y} = 0.0$. The Landau damping coefficient is
$\gamma = 0.1$. Increasing the damping coefficient $\gamma$ from
$0.1$ to $0.5$, the hump becomes broadened slightly as shown in
the inset figure.} \label{fig6}
\end{figure}

\subsection{Thermodynamics}

We find physical implication of the present study from
thermodynamics. Thermodynamics can be investigated in the
Luttinger-Ward functional approach \cite{LW}
\begin{widetext} \bqa && F[\boldsymbol{\Sigma}(k_0),\Pi(q_0,q)] =
- N T \sum_{k_{0}} \int \frac{d^{2}k}{(2\pi)^{2}} \boldsymbol{tr}
\Bigl\{ \ln \Bigl( \boldsymbol{\tau_{3}}
\boldsymbol{G}^{-1}(k_0,k) \Bigr) - \boldsymbol{\Sigma}(k_0)
\boldsymbol{\tau_{3}} \boldsymbol{G} (k_0,k) \Bigr\} \nn && + T
\sum_{q_0} \int \frac{d^{2}q}{(2\pi)^{2}} \Bigl\{ \ln \Bigl(
D^{-1}(q_0,q) \Bigr) - \Pi(q_0,q) D(q_0,q) \Bigr\} \nn && + T
\sum_{k_{0}} T \sum_{q_0} \int \frac{d^{2}k}{(2\pi)^{2}} \int
\frac{d^{2}q}{(2\pi)^{2}} \frac{1}{2} \boldsymbol{tr} \Bigl\{
\boldsymbol{\tau_{3}} \boldsymbol{G} (k_0+q_0,k+q) D(q_0,q)
\boldsymbol{\tau_{3}} \boldsymbol{G} (k_0,k) \Bigr\} , \eqa
\end{widetext} where the last term is called the Luttinger-Ward
functional, resulting from the Nambu-Eliashberg approximation in
this paper.

Using the Nambu-Eliashberg equation, one can simplify this
expression and obtain the following free energy \begin{widetext}
\bqa && F(T) = - T \sum_{k_{0}} \int_{-\infty}^{\infty}
\frac{dk_{y}}{2\pi} \int_{-\infty}^{\infty} \frac{dk_{x}}{2\pi}
\ln\Bigl\{ \Bigl( k_{y}^{2} + i \frac{\lambda }{N} \mbox{sgn}(k_0)
|k_0|^{\frac{2}{z}} \Bigr)^{2} - k_{x}^{2} -
\frac{\chi^{2}}{N^{2}} |k_0|^{\frac{4}{z}} \Bigr\} \nn && +
\frac{\Lambda^{2} }{2\pi} T \sum_{q_0} \int_{-\infty}^{\infty}
\frac{dq_{y}}{2\pi} \Bigl\{ \ln \Bigl( q_{y}^{z-1} + \gamma
\frac{|q_0|}{|q_y|} \Bigr) - \frac{\gamma \frac{|q_0|}{|q_y|}
}{q_{y}^{z-1} + \gamma \frac{|q_0|}{|q_y|}} \Bigr\} , \eqa
\end{widetext} where $\Lambda$ is a momentum cutoff.
It is trivial to observe that this free energy is qualitatively
the same as that from the simple Eliashberg approximation, given
by
\begin{widetext}\bqa && F(T) = - T \sum_{k_{0}} \int_{-\infty}^{\infty}
\frac{dk_{y}}{2\pi}  \int_{-\infty}^{\infty} \frac{dk_{x}}{2\pi}
\ln\Bigl\{ \Bigl( k_{y}^{2} + i \frac{\lambda }{N} \mbox{sgn}(k_0)
|k_0|^{\frac{2}{z}} \Bigr)^{2} - k_{x}^{2} \Bigr\} \nn && +
\frac{\Lambda^{2} }{2\pi} T \sum_{q_0} \int_{-\infty}^{\infty}
\frac{dq_{y}}{2\pi} \Bigl\{ \ln \Bigl( q_{y}^{z-1} + \gamma
\frac{|q_0|}{|q_y|} \Bigr) - \frac{\gamma \frac{|q_0|}{|q_y|}
}{q_{y}^{z-1} + \gamma \frac{|q_0|}{|q_y|}} \Bigr\} , \eqa
\end{widetext} except the anomalous self-energy part. In this
respect thermodynamics of the Nambu-Eliashberg theory is
essentially the same as that of the Eliashberg theory.

To perform integrals in Eq. (30) is straightforward. The boson
sector gives rise to the conventional $T^{2}$ contribution while
the fermion part results in the typical scaling expression $F(T)
\propto T^{(d+z)/z}$ with $z = 3$ and $d = 2$ in the low
temperature limit \cite{Kim_Adel_Pepin,Comment_Self_energy}.

An essential point in this demonstration is that thermodynamics is
determined by the $z = 3$ quantum criticality. This doe not seem
to be consistent with the previous perturbative analysis
\cite{Metlitski_Sachdev1} although this study did not examine the
thermodynamic potential in their perturbative scheme, thus a
direct comparison with the present result is not possible. To
construct the Luttinger-Ward functional consistent with this
perturbative scheme is another problem. But, we expect that the
singular part of the thermodynamic potential will be given by \bqa
&& F_{s}(T) \propto - \frac{1}{\beta} \sum_{k_{0}} \int
\frac{d^{2}{\boldsymbol{k}}}{(2\pi)^{2}} \ln [-G_{f}^{-1}(i
k_{0},\boldsymbol{k})] , \nonumber \eqa where $G_{f}(i
k_{0},\boldsymbol{k})$ is the full fermion propagator with the
AL-type vertex corrected self-energy. Since the fermion
self-energy has an anomalous exponent, the specific heat
coefficient will follow scaling different from $z = 3$.

One may claim that this difference is natural because the
Luttinger-Ward functional Eq. (30) does not seem to include the
AL-type vertex corrections. However, this is not true. The key
point in the present analysis is that the AL-type vertex
corrections are not only introduced but also summed up to an
infinite order in the Nambu-Eliashberg framework, taking into
account the anomalous pairing self-energy. As a result, the
anomalous exponent in the fermion self-energy disappears and
thermodynamics remains qualitatively the same as that of the
simple Eliashberg theory because the pairing self-energy is
essentially the same as the normal self-energy consistent with the
$z = 3$ scaling.

It should be pointed out that although we argued that the singular
contribution for such vertex corrections will be kept within this
anomalous pairing self-energy at least, we could not prove the
equivalence exactly and showed it approximately as discussed in
section II-B. If this "equivalence" turns out to be true, our
statement has important physical meaning. In this respect it will
be valuable to perform a direct summation for the AL-type vertex
corrections, expected to confirm our result.

\section{Discussion and perspectives}

\subsection{Discussion and summary}

Our study was motivated from a recent perturbative analysis
\cite{Metlitski_Sachdev1}, based on the Eliashberg theory as the
zeroth order, where the Aslamasov-Larkin (AL) -type vertex
corrections give rise to an anomalous novel exponent for the
fermion self-energy while they do not modify the Landau-damping
boson self-energy. If this is a general feature beyond this level
of approximation, various quantum critical phenomena \cite{HFQCP}
should be reconsidered because the emergence of the anomalous
exponent is expected to result in various novel critical exponents
for thermodynamics, transport, and etc. In this respect it was
crucial to investigate the existence of the anomalous exponent
beyond this level of approximation.

Unfortunately, it was not easy to perform a direct
non-perturbative summation for the AL-type vertex corrections due
to retardation effects of gauge fluctuations although one can
construct self-consistent equations analogous to superconducting
instability. Instead, hinted from superconductivity (section
II-B), we reformulated the non-perturbative summation, introducing
an anomalous self-energy in the Nambu-spinor formalism, where the
two components are momentum $k_{F}$ and $-k_{F}$ fermions,
respectively. As a result, we found that the anomalous exponent of
the normal self-energy turns out to disappear and the pairing
self-energy has essentially the same functional form as the
conventional Eliashberg self-energy for frequency. We argued that
this is the result from the non-perturbative summation for the
AL-type vertex corrections in a "unitary" transformed basis
(Nambu-spinor formalism).

One implication of this result could be found in thermodynamics.
If one evaluates the thermodynamic potential in a perturbative way
with the AL-type vertex corrections, the appearance of the
anomalous exponent in the fermion self-energy may give rise to a
novel exponent for temperature dependence in the specific heat
coefficient, different from the $z = 3$ scaling result. Here, $z$
is the dynamic critical exponent. However, we argued that the
non-perturbative summation for the AL-type vertex corrections,
performed in the Nambu-Eliashberg framework, leaves the specific
heat coefficient unchanged from the conventional Eliashberg theory
with $z = 3$. This is important for physical application to the
heavy fermion quantum critical point, particularly, for the Kondo
breakdown quantum critical point scenario \cite{KBQCP}, where the
$z = 3$ quantum criticality gives a qualitatively good description
for thermodynamics in a certain heavy fermion material
\cite{Kim_Adel_Pepin}.

More direct implication of the anomalous self-energy may be found
in spectroscopy measurements. For example, if the $2k_{F}$
susceptibility (spin or charge) is measured (or evaluated), we
expect that the anomalous propagator contributes and this will
show unexpected composite states from critical fermions ($k_{F}$
and $-k_{F}$) directly. Its detailed shape, for example, the peak
or hump-like position and the width (life time of the state), is
not clear at present. An optical conductivity (in the case of
nematic quantum criticality) will also show an interesting
feature. As expected, it would be enhanced at finite frequencies
due to breaking of such preformed particle-hole pairs.

We also discussed the role of 2$k_{F}$ ladder vertex corrections
in the critical state described by the simple Eliashberg theory.
In particular, the issue was to prove existence of such a critical
state as shows power law divergence of the 2$k_{F}$ spin
susceptibility while the fermion self-energy vanishes with a
nontrivial exponent. Although this physics is typical at quantum
criticality, we argued that this issue should be addressed more
sincerely in the critical spin liquid state, i.e.,
self-consistently even in the relativistic U(1) gauge theory,
where the $1/N$ expansion is well defined.

\subsection{Perspectives}

Although the Nambu-Eliashberg theory does not give an anomalous
exponent beyond the Eliashberg approximation, one can extend this
framework, introducing the ladder-type vertex correction
\begin{widetext}
\bqa && \Pi(q_0,q) = \boldsymbol{tr} \int \frac{d k_{0}}{2\pi}
\int \frac{d^{2} k}{(2\pi)^{2}}
\boldsymbol{\Lambda}(k_0+q_0,k+q;k_0,k) \boldsymbol{\tau_{3}}
\boldsymbol{G}(k_0+q_0,k+q) \boldsymbol{\tau_{3}}
\boldsymbol{G}(k_0,k) , \nn && \boldsymbol{\Sigma}(k_0) = -
\frac{1}{N} \int \frac{d q_{0}}{2\pi} \int \frac{d^{2}
q}{(2\pi)^{2}} \boldsymbol{\Lambda}(k_0+q_0,k+q;k_0,k)
\boldsymbol{\tau_{3}} \boldsymbol{G}(k_0+q_0,k+q) D(q_0,q) ,  \eqa
\end{widetext} where the two by two matrix vertex function is
given by the ladder approximation
\begin{widetext}
\bqa && \boldsymbol{\Lambda}(k_0+q_0,k+q;k_0,k) = \boldsymbol{I} -
\int \frac{d l_{0}}{2\pi} \int\frac{d^{2} l}{(2\pi)^{2}}
\boldsymbol{\Lambda}(k_0+q_0-l_0,k+q-l;k_0-l_0,k-l) \nn &&
\boldsymbol{\tau_{3}} \boldsymbol{G}_{\sigma}(k_0+q_0-l_0,k+q-l)
\boldsymbol{\tau_{3}} \boldsymbol{G}_{\sigma}(k_0-l_0,k-l)
D(l_0,l) . \eqa
\end{widetext} This can be regarded as natural generalization of
the one patch formulation, taking the Nambu representation to
incorporate 2$k_{F}$ backscattering. An important point is that
the vertex function satisfies the Ward identity automatically
because the Nambu-Eliashberg approximation is conserving, i.e.,
constructed from the Luttinger-Ward functional approach. We expect
that the same strategy as the one patch formulation can be applied
to the Nambu-Eliashberg theory for the two patch construction,
where the Ward identity will help us introduce such vertex
corrections \cite{Kim_Ladder}. Another advantage in this formalism
is that the 2$k_{F}$ ladder vertex will be introduced naturally.
It is an interesting future direction to solve these equations.

We would like to thank T. Takimoto and S.-S. Lee for helpful
discussions. K.-S. Kim was supported by the National Research
Foundation of Korea (NRF) grant funded by the Korea government
(MEST) (No. 2010-0074542).

\appendix

\section{Evaluation of the polarization function}

We prove that the Landau damping boson self-energy Eq. (25) is
preserved in the Nambu-Eliashberg theory. Inserting both normal
and abnormal fermion Green's functions into the equation of the
boson self-energy, we obtain
\begin{widetext}
\bqa && \Pi(q_0,q) = 2 \int_{-q_0}^{0} \frac{d k_{0}}{2\pi}
\int_{-\infty}^{\infty} \frac{d k_y}{2 \pi}
\int_{-\infty}^{\infty} \frac{d k_x}{2 \pi} \nn && \frac{ k_{x}
(k_{x} + q_{x}) + k_{y}^{2} (k_y+q_y)^{2} - i \frac{\lambda }{N}
|k_0 |^{\frac{2}{z}} k_{y}^{2} + i \frac{\lambda }{N}
|k_0+q_0|^{\frac{2}{z}} (k_y+q_y)^{2} + \frac{\lambda^{2} +
\chi^{2}}{N^{2}} |k_0+q_0|^{\frac{2}{z}} |k_0 |^{\frac{2}{z}} }{
\Bigl\{ (k_{x} + q_{x})^{2} - (k_{y}+q_{y})^{4} - 2 i
\frac{\lambda }{N} |k_0+q_0|^{\frac{2}{z}} (k_{y}+q_{y})^{2} +
\frac{\lambda^{2} + \chi^{2}}{N^{2}} |k_0+q_0|^{\frac{4}{z}}
\Bigr\} \Bigl\{ k_{x}^{2} - k_{y}^{4} + 2 i \frac{\lambda }{N}
|k_0 |^{\frac{2}{z}} k_{y}^{2} + \frac{\lambda^{2} + \chi^{2}
}{N^{2}} |k_0 |^{\frac{4}{z}} \Bigr\} } , \nn \eqa \end{widetext}
where both normal and pairing self-energies are explicitly used.

It is straightforward to perform the integration of $k_{x}$,
resulting in \begin{widetext} \bqa && \Pi(q_0,q) = \frac{1}{2\pi}
\int_{-q_0}^{0} \frac{d k_{0}}{2\pi} \int_{-\infty}^{\infty}
\frac{d k_y}{ 2\pi }\frac{ \sqrt{D}\Bigl((B+C)(C-D) +
(B-C)q_{x}^{2}\Bigr) + \sqrt{C}\Bigl((B+D)(C-D) -
(B-D)q_{x}^{2}\Bigr) }{ \sqrt{C}\sqrt{D}\Bigl((C-D)^{2} -
2(C+D)q_{x}^{2} + q_{x}^{4}\Bigr) } , \nn \eqa where \bqa B &=&
k_{y}^{2} (k_y+q_y)^{2} - i \frac{\lambda }{N} |k_0
|^{\frac{2}{z}} k_{y}^{2} + i \frac{\lambda }{N}
|k_0+q_0|^{\frac{2}{z}} (k_y+q_y)^{2} + \frac{\lambda^{2} +
\chi^{2}}{N^{2}} |k_0+q_0|^{\frac{2}{z}} |k_0 |^{\frac{2}{z}} ,
\nn C &=& (k_{y}+q_{y})^{4} + 2 i \frac{\lambda }{N}
|k_0+q_0|^{\frac{2}{z}} (k_{y}+q_{y})^{2} - \frac{\lambda^{2} +
\chi^{2}}{N^{2}} |k_0+q_0|^{\frac{4}{z}} , \nn D &=& k_{y}^{4} - 2
i \frac{\lambda }{N} |k_0 |^{\frac{2}{z}} k_{y}^{2} -
\frac{\lambda^{2} + \chi^{2} }{N^{2}} |k_0 |^{\frac{4}{z}} . \eqa
\end{widetext}

Scaling $k_{y}$ and $k_{0}$ as follows \bqa && k_{y} = q_{y} y ,
~~~~~ k_{0} = q_{0} t , \eqa the above expression reads
\begin{widetext}
\bqa && \Pi(q_0,q) = \frac{1}{2\pi} \frac{|q_{0}|}{ |q_{y}| }
\int_{-1}^{0} \frac{d t}{2\pi} \int_{-\infty}^{\infty} \frac{d y}{
2\pi } \frac{ \sqrt{d}\Bigl((b+c)(c-d) +
(b-c)\frac{q_{x}^{2}}{q_{y}^{2}}\Bigr) + \sqrt{c}\Bigl((b+d)(c-d)
- (b-d)\frac{q_{x}^{2}}{q_{y}^{2}}\Bigr) }{
\sqrt{c}\sqrt{d}\Bigl((c-d)^{2} -
2(c+d)\frac{q_{x}^{2}}{q_{y}^{2}} +
\frac{q_{x}^{4}}{q_{y}^{4}}\Bigr) } , \eqa where $b$, $c$, and $d$
are given by \bqa && B = q_{y}^{4} \Bigl\{ y^{2} (y+1)^{2} - i
\frac{\lambda }{N} |t|^{\frac{2}{z}} y^{2} \Bigl(
\frac{|q_{0}|^{\frac{1}{z}}}{|q_{y}| } \Bigr)^{2} + i
\frac{\lambda }{N} |t+1|^{\frac{2}{z}} (y+1)^{2} \Bigl(
\frac{|q_{0}|^{\frac{1}{z}}}{|q_{y}| } \Bigr)^{2} +
\frac{\lambda^{2} + \chi^{2}}{N^{2}} |t+1|^{\frac{2}{z}} |t
|^{\frac{2}{z}} \Bigl( \frac{|q_{0}|^{\frac{1}{z}}}{|q_{y}| }
\Bigr)^{4} \Bigr\} \equiv q_{y}^{4} b , \nn && C = q_{y}^{4}
\Bigl\{ (y+1)^{4} + 2 i \frac{\lambda }{N} |t+1|^{\frac{2}{z}}
(y+1)^{2} \Bigl( \frac{|q_{0}|^{\frac{1}{z}}}{|q_{y}| } \Bigr)^{2}
- \frac{\lambda^{2} + \chi^{2}}{N^{2}} |t+1|^{\frac{4}{z}} \Bigl(
\frac{|q_{0}|^{\frac{1}{z}}}{|q_{y}| } \Bigr)^{4} \Bigr\} \equiv
q_{y}^{4} c , \nn && D = q_{y}^{4} \Bigl\{y^{4} - 2 i
\frac{\lambda }{N} |t |^{\frac{2}{z}} y^{2} \Bigl(
\frac{|q_{0}|^{\frac{1}{z}}}{|q_{y}| } \Bigr)^{2} -
\frac{\lambda^{2} + \chi^{2} }{N^{2}} |t |^{\frac{4}{z}} \Bigl(
\frac{|q_{0}|^{\frac{1}{z}}}{|q_{y}|} \Bigr)^{4} \Bigr\} \equiv
q_{y}^{4} d . \eqa \end{widetext}

An important point is that this integral is convergent near $y
\approx 0$. As a result, the main contribution comes from \bqa &&
|y| \gg 1 . \eqa Scaling $y$ in the following way \bqa && Y = y
\Bigl( \frac{|q_{0}|^{\frac{1}{z}}}{|q_{y}| } \Bigr)^{-1} , \eqa
we obtain \bqa && c - d \approx 4 Y^{3} , ~~~~~ b - c \approx - 2
Y^{3} , ~~~~~ b - d \approx 2 Y^{3} \nn \eqa in the dominant
region for the integral. Finally, we find the boson self-energy
\begin{widetext}
\bqa && \Pi(q_0,q) \approx \frac{1}{ \pi} \frac{|q_{0}|}{ |q_{y}|
} \int_{-1}^{0} \frac{d t}{2\pi} \int_{\Bigl(
\frac{|q_{0}|^{\frac{1}{z}}}{|q_{y}|} \Bigr)^{-1}}^{\infty}
\frac{d Y}{ 2 \pi } \frac{ Y^{2} \Bigl\{4 Y^{7} - 2 Y^{3}
\frac{q_{x}^{2}}{q_{y}^{2}} \Bigl(
\frac{|q_{0}|^{\frac{1}{z}}}{|q_{y}|} \Bigr)^{-4}
\Bigr\}}{Y^{4}\Bigl\{16Y^{6} - 4 Y^{4}
\frac{q_{x}^{2}}{q_{y}^{2}}\Bigl(
\frac{|q_{0}|^{\frac{1}{z}}}{|q_{y}|} \Bigr)^{-2} +
\frac{q_{x}^{4}}{q_{y}^{4}} \Bigl(
\frac{|q_{0}|^{\frac{1}{z}}}{|q_{y}|} \Bigr)^{-6}\Bigr\}} \nn &&
\approx \frac{1}{ \pi} \frac{|q_{0}|}{ |q_{y}| } \int_{-1}^{0}
\frac{d t}{2\pi} \int_{\Bigl(
\frac{|q_{0}|^{\frac{1}{z}}}{|q_{y}|} \Bigr)^{-1}}^{\Lambda}
\frac{d Y}{ 2 \pi } \frac{1}{ 4Y } = \frac{1}{16\pi^{3}}
\frac{|q_{0}|}{ |q_{y}| } \Bigl\{ \ln \Lambda + \ln \Bigl(
\frac{|q_{0}|^{\frac{1}{z}}}{|q_{y}|} \Bigr) \Bigr\} \approx
\frac{ \ln \Lambda }{16\pi^{3}} \frac{|q_{0}|}{ |q_{y}| } , \eqa
\end{widetext} nothing but the Landau damping form. In this
respect Eq. (25) and Eq. (27) are the fully self-consistent
solution of the Nambu-Eliashberg equations.


\begin{thebibliography}{9}
\bibitem{Comment_Self_energy} This $z = 3$ scaling property can be understood from the total
self-energy in the normal Green's function, given by \bqa &&
\Sigma_{tot}^{s}(k_0,k) = \frac{|\Phi(k_0)|^{2}}{i \eta k_0 - s
k_{x} + k_{y}^{2} - \Sigma(k_0)} . \nonumber \eqa In the long wave
length limit this total self-energy can be approximated as follows
\bqa && \Sigma_{tot}^{s}(k_0,k) \approx - \frac{|\Phi(k_0)|^{2}}{
\Sigma(k_0)} = - i \frac{\chi^{2}/\lambda}{N} \mbox{sgn}(k_0)
|k_0|^{\frac{2}{z}} . \nonumber \eqa This expression is
essentially the same as the conventional Eliashberg self-energy.
In this respect we expect that low energy physics does not change
from the Eliashberg dynamics.
\end{thebibliography}
\end{document}